# Modeling of Periodic Array of Cut-through Slits with Periodic Surface Conductivity at the Interfaces of an Anisotropic Medium

Babak Rahmani and Khashayar Mehrany

The authors are with the Electrical Engineering Department, Sharif University of Technology, Tehran 11155-4363, Iran

Email: mehrany@sharif.edu

***Abstract***

A periodic arrangement of one-dimensional slits carved in perfect electric conductor is investigated and an equivalent model based on the effective medium theory is derived. The proposed model is no longer fully homogeneous and features periodic surface conductivity on its upper and lower interfaces. Therefore, and in sheer contrast to all the previous attempts that were successful in mimicking only the zeroth-order diffracted waves, it is capable of emulating both specular and non-specular diffraction orders. The parameters of the equivalent model are found by comparing the scattered waves of the proposed model against those of the original structure obtained by invoking the rigorous mode matching technique based on the single mode approximation inside the slits. The proposed model is characterized by diagonal anisotropic permittivity and permeability tensors together with a periodic surface conductivity at the interfaces of the metallic grating and the ambient environment. The accuracy of the model is verified by the full-wave simulations. The proposed approach might be useful for devising applications which work based on the characteristics of higher Floquet orders.

***Index Terms***

cut-through slit, diffraction order, effective medium model, extraordinary transmission, periodic surface conductivity.

## I. INTRODUCTION

PERIODIC metallic structures support a variety of unexpected electromagnetic features. Mimicking high refractive index materials [1]-[3], supporting enhanced transmission of light through sub-wavelength holes or slits perforated in opaque metallic films [4]-[7], which is usually referred to as the extraordinary transmission (EOT), and showing plasma-like behavior at microwave and terahertz (THz) frequencies [8] are among a few interesting phenomena that are brought about on account of sub-wavelength features incorporated in periodic metallic structures. The latter two above-mentioned aspects- EOT and plasma like behavior- are further discussed below since they cannot be always understood by using the standard effective medium theory even at normalized frequencies not larger than unity.

EOT in one- and two-dimensional metallic gratings can be of resonant or non-resonant nature. Thanks to their ultra-broadband characteristics, the non-resonant transparencies have been of much interest lately [9]. They appear because of the broadband impedance matching that might occur between the structure and its surrounding media at specific incident Brewster angles. The resonant EOT- on the other hand- is far more complex and can be either broad or sharp. The former type is caused by the Fabry-Perot (FP) resonance within every single slit of the structure while the latter is strongly affected by the collective grating resonance between adjacent slits. Since the FP resonance is the intrinsic property of each and every single slit in the structure, the broad resonance is normally independent of the incident angle and can be easily explained by invoking the effective medium theory [1]. Despite the simplicity of its nature, it finds miscellaneous applications from realization of transformation optical devices [10] to spectrum filters [11]. The sharp resonant EOT is Fano type and thus asymmetric. It is of more interest in sensing or filtering applications and cannot be understood without resorting to more complex models which include the effects of higher diffracted orders. It is a notable example of a physical phenomenon, which is observed at less than unity normalized frequency and yet cannot be explained by the conventional homogenization techniques.

The plasma-like behavior of periodic metallic structures at microwave and terahertz (THz) frequencies results in what is usually referred to as the spoof surface plasmon (SSP) oscillations [8] and permits new possibilities otherwise impossible to realize. One notable example is miniaturization of THz devices through high contrast grating-based SSPs [12]. Whereas the bound surface states associated with SSPs in structures whose metallic slits support electromagnetic modes with non-zero cut-off frequency can be delineated by their negative and thus plasma-like effective permittivity at all frequencies lying below the cut-off of the principal electromagnetic mode supported by the slit [8], the plasma-like behavior in other structures whose effective permittivity happens to be positive is rather mind boggling. For instance, the effective permittivity of the one-dimensional cut-through metallic slits whose principal transverse electromagnetic (TEM) mode has no cut-off frequency is always positive [1]. In such cases, the plasma-like behavior can be attributed to the folding of the band diagram rather than a negative effective permittivity and for that reason cannot be fully understood by the conventional homogenization techniques.

Miscellaneous analytical models with excellent accuracy for one- and two-dimensional metallic gratings have been proposed [13]-[19]. For instance, transmission line models that have been suggested for one-dimensional (1D) metallic gratings can successfully explicate non-resonant [14] and low frequency broad resonant EOTs [15], [16]. Neither the sharp resonant EOT, nor the plasma like behavior observed in 1D metallic grating can be fully understood by using the conventional models wherein the effect of non-specular diffraction orders is neglected. The first attempt toward explanation of sharp resonant EOT in 1D metallic structures was successfully made by including the effect of higher diffraction orders on the zeroth order diffracted wave in a shunt admittance added to the existent transmission line models [20]-[22]. In a similar fashion, an effective surface conductivity was added to the existent effective medium models to account for the effect of non-specular orders on the zeroth-order diffracted wave [23]-[25]. It is worth noting that even though the above-mentioned improved techniques can predict the sharp resonant EOT, they fail to duly explain the phenomenon in terms of the Fano resonance theory. Furthermore, they are incapable of fully explaining the plasma-like behavior when the effective permittivity is positive.

In this work, the existent effective medium theory comprised of effective electric permittivity and magnetic permeability tensors is modified by introducing a periodic surface conductivity upon its interfaces. Since the only periodic element of the model is its surface conductivity, the proposed model is much simpler than the original structure and yet has an excellent accuracy even at the normalized frequencies beyond unity. It can explain the plasma-like behavior in terms of the band bending caused by the periodicity of the surface conductivity and relate the sharp resonant EOT to the analytical expression given for Fano type asymmetric resonance [26], [27]. The proposed model gives the diffraction efficiency of the non-specular orders and thereby might find applications in structures designed to tailor the characteristics of higher Floquet orders such as beam-splitters [28].

The paper is organized as follows: In section II, the scattering parameters of a periodic arrangement of cut-through metallic slits are analytically calculated for both transverse magnetic (TM) and transverse electric (TE) incident waves. In section III, the detailed description of the proposed model is given for both major polarization states. In section IV, the parameters of the proposed model are extracted. Numerical examples are given in section V and finally the conclusions are drawn in section VI

## II. DERIVATION OF THE SCATTERING MATRIX IN CUT-THROUGH SLITS

In this section, we find the scattering parameters at the interface $z=0$ of the structure depicted in Fig. 1. The 1D grating is periodic with period *d* in the *x*-direction and it is uniform along the *y*-direction. The width of each slit is represented by *a*. For simplicity's sake, the surrounding media and the slit regions are assumed to be free space. Both major polarizations, i.e. TM and TE, are discussed in the following two subsections.

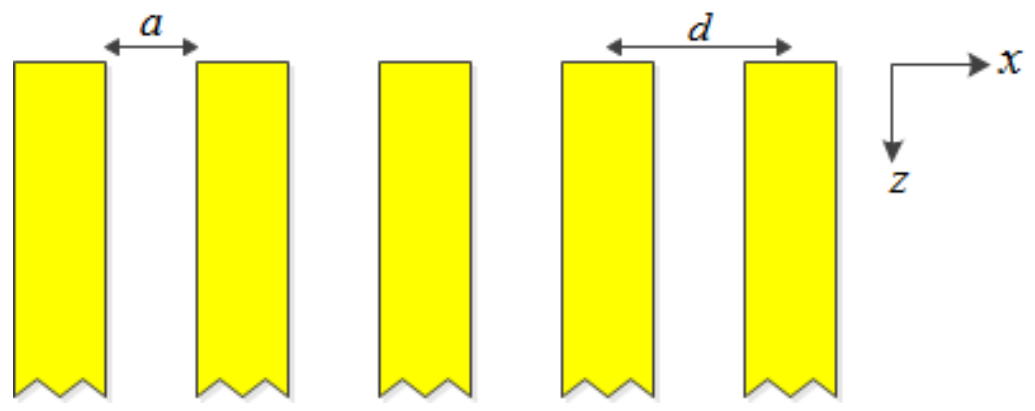

Fig. 1 Periodic arrangement of one-dimensional semi-infinite cut-through slits: $d$ is the period of the structure and $a$ is the width of each slit. Region I is the free space $z < 0$, and region II is the grating region $z > 0$.

### A. TM Polarization

The Bloch expansion of the TM polarized plane waves in region I can be written as follows

$$H_y^{(\mathrm{I})}(x,z) = \sum_{n=-\infty}^{\infty} (a_n e^{-jk_{zn}z} + r_n e^{jk_{zn}z}) e^{-jk_{xn}x},$$

$$E_x^{(\mathrm{I})}(x,z) = \sum_{n=-\infty}^{\infty} \xi_{1n} (a_n e^{-jk_{zn}z} - r_n e^{jk_{zn}z}) e^{-jk_{xn}x} \tag{1}$$

where

$$k_{xn} = k_{x0} + \frac{2n\pi}{d}, \quad (n = 0, \pm1, \pm2, \ldots) \tag{2}$$

$$k_{zn} = \sqrt{k_0^2 - k_{xn}^2}, \quad (n = 0, \pm1, \pm2, \ldots) \tag{3}$$

and

$$\xi_{1n} = \frac{k_{zn}\eta_0}{k_0} \tag{4}$$

In these expressions, $k_{x0}$ is the Bloch wavenumber and can be related to the angle of incidence $\theta$ by $k_{x0} = k_0 \sin\theta$, the subscript $n$ represents the diffraction order, $k_0$ represents the free space wavenumber and $\eta_0$ stands for the intrinsic impedance of the free space. It is worth noting that $\xi_{1n}$ is in fact the TM impedance of the $n$th diffraction order in the free space whose propagation constant along the $z$-direction, i.e. $k_{zn}$, is defined in such a manner that it is either a positive real number corresponding to a propagating wave or a negative imaginary number corresponding to an evanescent wave. Obviously, the Bloch expansion with infinite terms has to be inevitably truncated by keeping $2N + 1$ diffraction orders, where $-N \le n \le N$.

The transverse components of the corresponding electromagnetic waves for $0<x<a$ in region II can be approximated by

$$H_y^{(\mathrm{II})}(x,z) = t_0 e^{-j\beta_0 z} + b_0 e^{+j\beta_0 z},$$

$$E_x^{(\mathrm{II})}(x,z) = \eta_0 (t_0 e^{-j\beta_0 z} - b_0 e^{+j\beta_0 z}) \tag{5}$$

where $\beta_0$ is the propagation constant of the principal TEM electromagnetic mode supported by the slits which reads as

$$\beta_0 = k_0 \tag{6}$$

Here, the higher order modes are assumed to be below the cut-off and neglected. This expression yields accurate enough answers where the angular frequency is below $\omega < \pi c / a$ in which $c$ is the light's speed in the free space [16].

The sought after scattering parameters can be found by applying the continuity condition of tangential electromagnetic fields at the interface between regions I and II, i.e. at $z = 0$. First, the continuity condition of tangential electric fields at every point within the unit cell of the structure yields the following 2$N$+1 equations when the continuity condition of tangential electric fields is multiplied by $e^{jk_{xn}x}$ before being integrated over one period:

$$\boldsymbol{Z}^{(\mathrm{I})}(\boldsymbol{A}-\boldsymbol{R}) = \boldsymbol{P}\eta_0\kappa(t_0 - b_0) \tag{7}$$

Second, the continuity condition of tangential magnetic fields yields the following single equation when it is integrated over the slit width

$$(\boldsymbol{P}^*)^t(\boldsymbol{A}+\boldsymbol{R}) = \kappa(t_0 + b_0) \tag{8}$$

In these expressions, $\kappa = \sqrt{a/d}$, $\boldsymbol{A} = [a_n]_{(2N+1)\times 1}$, $\boldsymbol{R} = [r_n]_{(2N+1)\times 1}$, and $\boldsymbol{Z}^{(\mathrm{I})}$ is the TM impedance matrix of the region I given by:

$$\boldsymbol{Z}^{(\mathrm{I})} = [z_{ij}^{(\mathrm{I})}]_{(2N+1)\times(2N+1)},\; z_{ij}^{(\mathrm{I})} = \xi_{1(i-N-1)}\delta_{ij} \tag{9}$$

where $\delta_{ij}$ is the Kronecker delta function. Furthermore, $\boldsymbol{P}$ is the coupling vector and is written as

$$\boldsymbol{P} = [p_n]_{(2N+1)\times 1},\; p_n = p_{(n-N-1)}^{+} \tag{10}$$

where

$$p_n^{+} = \frac{1}{\sqrt{ad}}\int_0^a e^{+jk_{xn}x}dx \tag{11}$$

Given that the scattering parameters relate $\boldsymbol{R}$ and $t_0$ to $\boldsymbol{A}$ and $b_0$ via the following equations

$$\begin{bmatrix}\boldsymbol{R}\\ t_0\end{bmatrix} = [S]\begin{bmatrix}\boldsymbol{A}\\ b_0\end{bmatrix}, \tag{12$a$}$$

$$[S]_{(2N+2)\times(2N+2)} = \begin{bmatrix}\boldsymbol{S}_{11} & \boldsymbol{S}_{12}\\ \boldsymbol{S}_{21} & \boldsymbol{S}_{22}\end{bmatrix} \tag{12$b$}$$

they can be easily obtained by algebraic manipulations of (7) and (8):

$$[\boldsymbol{S}_{11}]_{(2N+1)\times(2N+1)} = (\boldsymbol{G}\kappa^{-1}(\boldsymbol{P}^*)^t + \boldsymbol{Z}^{(\mathrm{I})})^{-1}(-\boldsymbol{G}\kappa^{-1}(\boldsymbol{P}^*)^t + \boldsymbol{Z}^{(\mathrm{I})}) \tag{13$a$}$$

$$[\boldsymbol{S}_{12}]_{(2N+1)\times 1} = 2(\boldsymbol{G}\kappa^{-1}(\boldsymbol{P}^*)^t + \boldsymbol{Z}^{(\mathrm{I})})^{-1}\boldsymbol{G} \tag{13$b$}$$

$$[\boldsymbol{S}_{21}]_{1\times(2N+1)} = 2(\kappa + (\boldsymbol{P}^*)^t\boldsymbol{Z}^{(\mathrm{I})^{-1}}\boldsymbol{G})^{-1}(\boldsymbol{P}^*)^t \tag{13$c$}$$

$$[\boldsymbol{S}_{22}]_{1\times 1} = (\kappa + (\boldsymbol{P}^*)^t\boldsymbol{Z}^{(\mathrm{I})^{-1}}\boldsymbol{G})^{-1}(-\kappa + (\boldsymbol{P}^*)^t\boldsymbol{Z}^{(\mathrm{I})^{-1}}\boldsymbol{G}) \tag{13$d$}$$

where $\boldsymbol{G} = \boldsymbol{P}\eta_0\kappa$.

It is worth noting that once the scattering parameters are found and thus $t_0$ is duly related to $\boldsymbol{A}$ and $b_0$, the electromagnetic fields in the entire region II can be easily obtained by using the Bloch wave expansion of the electromagnetic fields within the unit cell of the structure, which are given in (6) for $0<x<a$ and are zero otherwise:

$$H_y^{(\mathrm{II})}(x,z) = \sum_{n=-\infty}^{\infty}(\tilde{t}_n e^{-j\beta_0 z} + \tilde{b}_n e^{j\beta_0 z})e^{-jk_{xn}x},$$

$$E_x^{(\mathrm{II})}(x,z) = \sum_{n=-\infty}^{\infty}\eta_0(\tilde{t}_n e^{-j\beta_0 z} - \tilde{b}_n e^{j\beta_0 z})e^{-jk_{xn}x} \tag{14}$$

where $\tilde{t}_n = p_n^+\kappa t_0$ and $\tilde{b}_n = p_n^+\kappa b_0$. The mathematical expression of tangential electromagnetic fields is needed to retrieve the parameters of the effective medium model in the next sections.

*B. TE Polarization*

In this part, we follow a similar procedure to obtain the scattering parameters of the structure for the TE polarization. The Bloch expansion of the TE polarized plane waves in region I can be written as follows

$$E_y^{(\mathrm{I})}(x,z) = \sum_{n=-\infty}^{\infty}(a_n e^{-jk_{zn}z} + r_n e^{jk_{zn}z})e^{-jk_{xn}x},$$

$$H_x^{(\mathrm{I})}(x,z) = \sum_{n=-\infty}^{\infty}\varsigma_{1n}(-a_n e^{-jk_{zn}z} + r_n e^{jk_{zn}z})\,e^{-jk_{xn}x} \tag{15}$$

where

$$\varsigma_{1n} = \frac{k_{zn}}{k_0\eta_0} \tag{16}$$

is the TE admittance of the $n$th diffraction order in the free space. Once again, the Bloch expansion with infinite terms is truncated and only $2N+1$ diffraction orders are kept.

Based on the single mode approximation within each slit, the transverse components of the electromagnetic waves for the interval $0<x<a$ in region II can be written in as

$$E_y^{(\mathrm{II})}(x,z) = \sin(\frac{\pi x}{a})(t_1 e^{-j\beta_1 z} + b_1 e^{+j\beta_1 z}),$$

$$H_x^{(\mathrm{II})}(x,z) = \sin(\frac{\pi x}{a})\frac{\beta_1}{k_0\eta_0}(-t_1 e^{-j\beta_1 z} + b_1 e^{+j\beta_1 z}) \tag{17}$$

where $\beta_1$ is the propagation constant of the principal TE mode supported by the slits which reads as

$$\beta_1 = \sqrt{k_0^2 - (\frac{\pi}{a})^2} \tag{18}$$

It is worth noting that $\beta_1$ can be either a positive real number corresponding to a propagating wave or a negative imaginary number corresponding to an evanescent wave. The single mode approximation is valid as long as the angular frequency is below the cut-off frequency of the second higher order TE mode, i.e. $\omega < 2\pi c/a$ [16].

In a similar fashion, the continuity condition of tangential electromagnetic fields at the interface $z=0$ is applied to obtain the sought-after scattering parameters. The continuity condition of tangential electric fields at every point within the unit cell of the structure yields the following $2N+1$ equations when the continuity of tangential electric fields is multiplied by $e^{jk_{xn}x}$ before being integrated over one period:

$$(\boldsymbol{A}+\boldsymbol{R}) = \boldsymbol{Q}\kappa(t_1+b_1) \tag{19}$$

The continuity condition of tangential magnetic fields leads to the following single equation once it is multiplied by $\sin(\pi x/a)$ and then integrated over one slit width

$$(\boldsymbol{Q}^*)^t \boldsymbol{Y}^{(\mathrm{I})}(\boldsymbol{A}-\boldsymbol{R}) = \frac{\beta_1 \kappa}{2k_0\eta_0}(t_1 - b_1) \tag{20}$$

In these expressions, $\boldsymbol{Y}^{(\mathrm{I})}$ is the TE admittance matrix of the region I given by:

$$\boldsymbol{Y}^{(\mathrm{I})} = [y_{ij}^{(\mathrm{I})}]_{(2N+1)\times(2N+1)},\ y_{ij}^{(\mathrm{I})} = \varsigma_{1(i-N-1)}\delta_{ij} \tag{21}$$

and $\boldsymbol{Q}$ is the coupling vector given by:

$$\boldsymbol{Q} = [q_n]_{(2N+1)\times 1},\ q_n = q_{(i-N-1)}^{+} \tag{22}$$

where

$$q_n^{+} = \frac{1}{\sqrt{ad}}\int_0^a \sin(\frac{\pi x}{a}) e^{+jk_{xn}x} dx \tag{23}$$

The scattering parameters can be easily obtained by algebraic manipulations of (19) and (20):

$$[\boldsymbol{S}_{11}]_{(2N+1)\times(2N+1)} = (\boldsymbol{Q}\kappa\boldsymbol{W}V^{-1} + \boldsymbol{I})^{-1}(\boldsymbol{Q}\kappa\boldsymbol{W}V^{-1} - \boldsymbol{I}) \tag{24a}$$

$$[\boldsymbol{S}_{12}]_{(2N+1)\times 1} = 2(\boldsymbol{Q}\kappa\boldsymbol{W}V^{-1} + \boldsymbol{I})^{-1}\boldsymbol{Q}\kappa \tag{24b}$$

$$[\boldsymbol{S}_{21}]_{1\times(2N+1)} = 2(V + \boldsymbol{W}\boldsymbol{Q}\kappa)^{-1}\boldsymbol{W} \tag{24c}$$

$$[\boldsymbol{S}_{22}]_{1\times 1} = (V + \boldsymbol{W}\boldsymbol{Q}\kappa)^{-1}(V - \boldsymbol{W}\boldsymbol{Q}\kappa) \tag{24d}$$

where $\boldsymbol{W} = (\boldsymbol{Q}^*)^t \boldsymbol{Y}^{(\mathrm{I})}$ and $V = \beta_1\kappa / 2k_0\eta_0$.

Much like the previous sub-section, the TE polarized electromagnetic fields in the entire region II can be easily obtained by using the Bloch wave expansion of the electromagnetic fields within the unit cell of the structure, which are given in (17) for 0<$x$<$a$ and are zero otherwise:

$$E_y^{(\mathrm{II})}(x,z) = \sum_{n=-\infty}^{\infty}(\tilde{t}_n e^{-j\beta_1 z} + \tilde{b}_n e^{j\beta_1 z})e^{-jk_{xn}x},$$
$$H_x^{(\mathrm{II})}(x,z) = \sum_{n=-\infty}^{\infty}\frac{\beta_1}{k_0\eta_0}(-\tilde{t}_n e^{-j\beta_1 z} + \tilde{b}_n e^{j\beta_1 z})e^{-jk_{xn}x} \tag{25}$$

where $\tilde{t}_n = q_n^{+}\kappa t_1$ and $\tilde{b}_n = q_n^{+}\kappa b_1$.

## III. Scattering in the Effective Medium Model

The effective medium model which is schematically shown in Fig. 2 is proposed for the 1D semi-infinite metallic grating depicted in Fig. 1. The model is comprised of an anisotropic medium whose electric permittivity and magnetic permeability tensors are homogenous and denoted by $\overline{\overline{\varepsilon}}_r$ and $\overline{\overline{\mu}}_r$, respectively. Modeling the semi-infinite metallic grating with homogeneous anisotropic medium has been already reported in the literature [2], [8], [23]. In our proposed model, the interface between the anisotropic medium and the free space has an inhomogeneous periodic anisotropic surface conductivity whose $x$ and $y$ components are $\sigma_x$ and $\sigma_y$ affecting the TM and the TE polarized waves, respectively.

In the following subsections, the scattering parameters of the proposed model are rigorously extracted for both major polarizations.

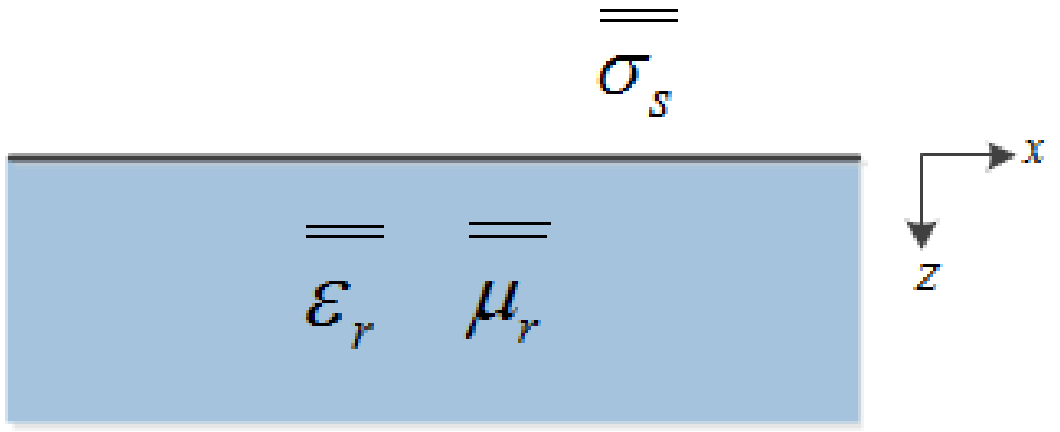

Fig. 2 Effective medium model for periodic arrangement of one-dimensional semi-infinite cut-through slits. Region I is the free space $z < 0$, and region II is the anisotropic region $z > 0$. There is a periodic surface conductivity at the interface $z = 0$.

### *A. TM Polarization*

Since the proposed model is periodic, the Bloch expansion of the TM polarized plane waves in the free space (region I) can be written as follows

$$\bar{H}_y^{(\mathrm{I})}(x,z) = \sum_{n=-\infty}^{\infty} (\bar{a}_n e^{-jk_{zn}z} + \bar{r}_n e^{jk_{zn}z}) e^{-jk_{xn}x},$$

$$\bar{E}_x^{(\mathrm{I})}(x,z) = \sum_{n=-\infty}^{\infty} \xi_{1n} (\bar{a}_n e^{-jk_{zn}z} - \bar{r}_n e^{jk_{zn}z}) e^{-jk_{xn}x} \tag{26}$$

Obviously, the electromagnetic fields in these expressions are essentially the same as those in (1).

In the anisotropic medium (region II), the Bloch wave expansion for the TM polarized waves can be written as follows

$$\bar{H}_y^{(\mathrm{II})}(x,z) = \sum_{n=-\infty}^{\infty} (\bar{t}_n e^{-j\bar{k}_{zn}^{\mathrm{TM}}z} + \bar{b}_n e^{j\bar{k}_{zn}^{\mathrm{TM}}z}) e^{-jk_{xn}x},$$

$$\bar{E}_x^{(\mathrm{II})}(x,z) = \sum_{n=-\infty}^{\infty} \bar{\xi}_{2n} (\bar{t}_n e^{-j\bar{k}_{zn}^{\mathrm{TM}}z} - \bar{b}_n e^{j\bar{k}_{zn}^{\mathrm{TM}}z}) e^{-jk_{xn}x} \tag{27}$$

where

$$\bar{k}_{zn}^{\mathrm{TM}} = \sqrt{k_0^2 \varepsilon_x \mu_y - \frac{\varepsilon_x}{\varepsilon_z} k_{xn}^2} \tag{28}$$

is the propagation constant along the $z$-direction, and

$$\bar{\xi}_{2n} = \frac{\bar{k}_{zn}^{\mathrm{TM}} \eta_0}{k_0 \varepsilon_x} \tag{29}$$

is the TM impedance of the $n$th diffraction order in the anisotropic region. It should be noted that $\bar{k}_{zn}^{\mathrm{TM}}$ can be either a positive real or a negative imaginary number. The former corresponds to a propagating wave while the latter corresponds to an evanescent wave.

By applying the standard boundary conditions at the interface of the two regions, the unknown coefficients in (26) and (27) can be determined when the Bloch expansions in both regions I and II are truncated by keeping only $2N$+1 diffraction orders. Applying the continuity condition of tangential electric fields across the interface $z$=0 yields the following $2N$+1 equations which are obtained by multiplying the continuity condition by $e^{jk_{xn}x}$ and integrating over one period

$$\boldsymbol{Z}^{(\mathrm{I})}(\bar{\boldsymbol{A}} - \bar{\boldsymbol{R}}) = \bar{\boldsymbol{Z}}^{(\mathrm{II})}(\bar{\boldsymbol{T}} - \bar{\boldsymbol{B}}) \tag{30}$$

The discontinuity condition of tangential magnetic fields across the interface z=0 leads to the following $2N$+1 equations when it is multiplied by $e^{jk_{xn}x}$ and then is integrated over one period

$$(\bar{\boldsymbol{A}} + \bar{\boldsymbol{R}}) = (\boldsymbol{I} + \tilde{\boldsymbol{\sigma}}_x \bar{\boldsymbol{Z}}^{(\mathrm{II})})\bar{\boldsymbol{T}} + (\boldsymbol{I} - \tilde{\boldsymbol{\sigma}}_x \bar{\boldsymbol{Z}}^{(\mathrm{II})})\bar{\boldsymbol{B}} \tag{31}$$

In these expressions, $\bar{\boldsymbol{A}} = [\bar{a}_n]_{(2N+1)\times 1}$, $\bar{\boldsymbol{B}} = [\bar{b}_n]_{(2N+1)\times 1}$, $\bar{\boldsymbol{T}} = [\bar{t}_n]_{(2N+1)\times 1}$, $\bar{\boldsymbol{R}} = [\bar{r}_n]_{(2N+1)\times 1}$, $\boldsymbol{Z}^{(\mathrm{I})}$ is already given in (9) while $\bar{\boldsymbol{Z}}^{(\mathrm{II})}$ is the TM impedance matrix of region II given by:

$$\bar{\boldsymbol{Z}}^{(\mathrm{II})} = [\bar{z}_{ij}^{(\mathrm{II})}]_{(2N+1)\times(2N+1)},\ \bar{z}_{ij}^{(\mathrm{II})} = \bar{\xi}_{2(i-N-1)}\delta_{ij} \tag{32}$$

Moreover, $I$ is a $(2N+1)\times(2N+1)$ identity matrix and $\tilde{\boldsymbol{\sigma}}_x$ is a matrix given by:

$$\tilde{\boldsymbol{\sigma}}_x = [\tilde{\sigma}_{ij}]_{(2N+1)\times(2N+1)},\ \tilde{\sigma}_{ij} = \begin{cases} \sigma_x^{j-i} & |j-i| \le N \\ 0 & o.w. \end{cases} \tag{33}$$

where $\sigma_x^{j-i}$ is the $(j-i)$ th Fourier expansion coefficient of the periodic $\sigma_x$.

The scattering parameters relate $\bar{\boldsymbol{R}}$ and $\bar{\boldsymbol{T}}$ to $\bar{\boldsymbol{A}}$ and $\bar{\boldsymbol{B}}$ via the following equations

$$\begin{bmatrix} \bar{\boldsymbol{R}} \\ \bar{\boldsymbol{T}} \end{bmatrix} = [\bar{\boldsymbol{S}}] \begin{bmatrix} \bar{\boldsymbol{A}} \\ \bar{\boldsymbol{B}} \end{bmatrix}, \tag{34$a$}$$

$$[\bar{\boldsymbol{S}}]_{(4N+2)\times(4N+2)} = \begin{bmatrix} \bar{\boldsymbol{S}}_{11} & \bar{\boldsymbol{S}}_{12} \\ \bar{\boldsymbol{S}}_{21} & \bar{\boldsymbol{S}}_{22} \end{bmatrix} \tag{34$b$}$$

and they can be easily obtained by algebraic manipulations of (30) and (31):

$$[\bar{\boldsymbol{S}}_{11}]_{(2N+1)\times(2N+1)} = (\boldsymbol{\chi} + \tilde{\boldsymbol{\sigma}}_x \bar{\boldsymbol{Z}}^{(\mathrm{II})}\boldsymbol{\chi} + \boldsymbol{I})^{-1}(\boldsymbol{\chi} + \tilde{\boldsymbol{\sigma}}_x \bar{\boldsymbol{Z}}^{(\mathrm{II})}\boldsymbol{\chi} - \boldsymbol{I}) \tag{35$a$}$$

$$[\bar{\boldsymbol{S}}_{12}]_{(2N+1)\times(2N+1)} = 2(\boldsymbol{\chi} + \tilde{\boldsymbol{\sigma}}_x \bar{\boldsymbol{Z}}^{(\mathrm{II})}\boldsymbol{\chi} + \boldsymbol{I})^{-1} \tag{35$b$}$$

$$[\bar{\boldsymbol{S}}_{21}]_{(2N+1)\times(2N+1)} = 2(\tilde{\boldsymbol{\sigma}}_x \bar{\boldsymbol{Z}}^{(\mathrm{II})} + \boldsymbol{\chi}^{-1} + \boldsymbol{I})^{-1} \tag{35$c$}$$

$$[\bar{\boldsymbol{S}}_{22}]_{(2N+1)\times(2N+1)} = (\tilde{\boldsymbol{\sigma}}_x \bar{\boldsymbol{Z}}^{(\mathrm{II})} + \boldsymbol{\chi}^{-1} + \boldsymbol{I})^{-1}(\tilde{\boldsymbol{\sigma}}_x \bar{\boldsymbol{Z}}^{(\mathrm{II})} + \boldsymbol{\chi}^{-1} - \boldsymbol{I}) \tag{35$d$}$$

where $\boldsymbol{\chi} = (\bar{\boldsymbol{Z}}^{(\mathrm{II})})^{-1}\boldsymbol{Z}^{(\mathrm{I})}$.

## *B. TE Polarization*

In a similar fashion, the Bloch expansion of the TE polarized plane waves in region I is written as follows:

$$\bar{E}_y^{(\mathrm{I})}(x,z) = \sum_{n=-\infty}^{\infty} (\bar{a}_n e^{-jk_{zn}z} + \bar{r}_n e^{jk_{zn}z}) e^{-jk_{xn}x},$$

$$\bar{H}_x^{(\mathrm{I})}(x,z) = \sum_{n=-\infty}^{\infty} \varsigma_{1n}(-\bar{a}_n e^{-jk_{zn}z} + \bar{r}_n e^{jk_{zn}z}) e^{-jk_{xn}x} \tag{36}$$

The electromagnetic fields in these expressions are essentially the same as those in (15).

The Bloch wave expansion for the TE polarized waves in region II can be written as follows

$$\bar{E}_y^{(\mathrm{II})}(x,z) = \sum_{n=-\infty}^{\infty} (\bar{t}_n e^{-j\bar{k}_{zn}^{\mathrm{TE}}z} + \bar{b}_n e^{j\bar{k}_{zn}^{\mathrm{TE}}z}) e^{-jk_{xn}x},$$

$$\bar{H}_x^{(\mathrm{II})}(x,z) = \sum_{n=-\infty}^{\infty} \bar{\varsigma}_{2n}(-\bar{t}_n e^{-j\bar{k}_{zn}^{\mathrm{TE}}z} + \bar{b}_n e^{j\bar{k}_{zn}^{\mathrm{TE}}z}) e^{-jk_{xn}x} \tag{37}$$

where

$$\bar{k}_{zn}^{\mathrm{TE}} = \sqrt{k_0^2 \varepsilon_y \mu_x - \frac{\mu_x}{\mu_z} k_{xn}^2} \tag{38}$$

is the propagation constant along the $z$-direction and

$$\bar{\varsigma}_{2n} = \frac{\bar{k}_{zn}^{\mathrm{TE}}}{k_0 \eta_0 \mu_x} \tag{39}$$

is the TE admittance of the $n$th diffraction order in the anisotropic region. Once again, $\bar{k}_{zn}^{\mathrm{TE}}$ is either a positive real (propagating waves) or a negative imaginary (evanescent waves) number. Inevitably, only (2$N$+1) orders are kept in the Bloch expansion in both regions I and II.

The sought-after scattering matrix can be obtained by applying the standard boundary conditions at the interface of the two regions. The continuity of tangential electric fields across the interface at z=0 results in the following 2$N$+1 equations when the continuity condition is multiplied by $e^{jk_{xn}x}$ before being integrated over one period

$$\overline{\boldsymbol{A}} + \overline{\boldsymbol{R}} = \overline{\boldsymbol{T}} + \overline{\boldsymbol{B}} \tag{40}$$

The discontinuity of tangential magnetic fields across the interface at $z = 0$ necessitates the following 2$N$+1 equations when the discontinuity condition is first multiplied by $e^{jk_{xn}x}$ and then is integrated over one period

$$\boldsymbol{Y}^{(\mathrm{I})}(\overline{\boldsymbol{A}} - \overline{\boldsymbol{R}}) = (\tilde{\boldsymbol{\sigma}}_y + \overline{\boldsymbol{Y}}^{(\mathrm{II})})\overline{\boldsymbol{T}} + (\tilde{\boldsymbol{\sigma}}_y - \overline{\boldsymbol{Y}}^{(\mathrm{II})})\overline{\boldsymbol{B}} \tag{41}$$

Here, $\boldsymbol{Y}^{(\mathrm{I})}$ was already given in (21) while $\overline{\boldsymbol{Y}}^{(\mathrm{II})}$ is in fact the TE admittance matrix of the region II and can be written as

$$\overline{\boldsymbol{Y}}^{(\mathrm{II})} = [\bar{y}_{ij}^{(\mathrm{II})}]_{(2N+1)\times(2N+1)},\ \bar{y}_{ij}^{(\mathrm{II})} = \bar{\varsigma}_{2(i-N-1)}\delta_{ij} \tag{42}$$

where $\tilde{\boldsymbol{\sigma}}_y$ is a matrix given by:

$$\tilde{\boldsymbol{\sigma}}_y = [\tilde{\sigma}_{ij}]_{(2N+1)\times(2N+1)},\ \tilde{\sigma}_{ij} = \begin{cases} \sigma_y^{j-i} & |j-i| \le N \\ 0 & o.w. \end{cases} \tag{43}$$

where $\sigma_y^{j-i}$ is the $(j-i)$th Fourier expansion coefficient of the periodic $\sigma_y$. The scattering parameters can be easily obtained by algebraic manipulations of (40) and (41):

$$\left[\overline{\boldsymbol{S}}_{11}\right]_{(2N+1)\times(2N+1)} = (\boldsymbol{Y}^{(\mathrm{I})} + \tilde{\boldsymbol{\sigma}}_y + \overline{\boldsymbol{Y}}^{(\mathrm{II})})^{-1}(\boldsymbol{Y}^{(\mathrm{I})} - \tilde{\boldsymbol{\sigma}}_y - \overline{\boldsymbol{Y}}^{(\mathrm{II})}) \tag{44a}$$

$$\left[\overline{\boldsymbol{S}}_{12}\right]_{(2N+1)\times(2N+1)} = 2(\boldsymbol{Y}^{(\mathrm{I})} + \tilde{\boldsymbol{\sigma}}_y + \overline{\boldsymbol{Y}}^{(\mathrm{II})})^{-1}\overline{\boldsymbol{Y}}^{(\mathrm{II})} \tag{44b}$$

$$\left[\overline{\boldsymbol{S}}_{21}\right]_{(2N+1)\times(2N+1)} = 2(\overline{\boldsymbol{Y}}^{(\mathrm{II})} + \tilde{\boldsymbol{\sigma}}_y + \boldsymbol{Y}^{(\mathrm{I})})^{-1}\boldsymbol{Y}^{(\mathrm{I})} \tag{44c}$$

$$\left[\overline{\boldsymbol{S}}_{22}\right]_{(2N+1)\times(2N+1)} = (\overline{\boldsymbol{Y}}^{(\mathrm{II})} + \tilde{\boldsymbol{\sigma}}_y + \boldsymbol{Y}^{(\mathrm{I})})^{-1}(\overline{\boldsymbol{Y}}^{(\mathrm{II})} - \tilde{\boldsymbol{\sigma}}_y - \boldsymbol{Y}^{(\mathrm{I})}) \tag{44d}$$

## IV. Extraction of the Bulk and Surface Parameters in the Effective Medium Model

One might presume that the parameters of the proposed model, i.e. components of the matrices $\overline{\overline{\varepsilon}}_r$ and $\overline{\overline{\mu}}_r$ as well as the Fourier coefficients of the periodic $\sigma_x$ and $\sigma_y$, can be easily obtained by comparing the scattering matrix of the original structure, i.e. (13) and (24), against that of the effective medium model, i.e. (35) and (44). The most obvious obstacle is that the scattering matrices of the original structure and that of the effective medium model are of different sizes. The former is a $(2N+2)\times(2N+2)$ matrix while the latter is $(4N+2)\times(4N+2)$. However, there are (2$N$+1) scattering parameters $\boldsymbol{S}_{11}$ and $\bar{\boldsymbol{S}}_{11}$ for both the original structure and the proposed model. Therefore, one might be tempted to compare $\boldsymbol{S}_{11}$ and $\bar{\boldsymbol{S}}_{11}$ against each other to extract the yet unknown parameters of the proposed model. That would be a futile attempt since the comparison fails to provide the $\tilde{\boldsymbol{\sigma}}_x$ and $\tilde{\boldsymbol{\sigma}}_y$ matrices.

The difficulty of finding the appropriate parameters for the proposed model which could accurately predict the scattering of the wave impinging upon the original structure originates from the fact that the electromagnetic expressions in region II of the original structure and that of the effective medium model differ from each other (see (5) and (27) for the TM polarization, and (17) and (37) for the TE polarization) even though the electromagnetic expressions in region I, i.e. the free space, of both the original structure and the effective medium model are essentially the same (see (1) and (26) for the TM polarization, and (15) and (36) for the TE polarization). In other words, one can easily compare $\bar{\boldsymbol{R}}$ against $\boldsymbol{R}$ but cannot compare $\bar{\boldsymbol{T}}$ against $t_0$ in a straightforward fashion. This difficulty can be overcome by noting that the electromagnetic fields in region II of the original structure, i.e. the slit region, can also be expanded in accordance with the Bloch theorem to resemble the electromagnetic expressions in region II of the proposed model. Comparison between (14) and (27) for the TM polarization and (25) and (37) for the TE polarization clearly demonstrates the strong resemblance. This approach is adopted in the following subsections and the parameters of the model are successfully extracted.

*A. TM Polarization*

Bloch expansion of the TM polarized electromagnetic fields expressions in region II of the original structure given in (14), results in $\boldsymbol{T}=[\tilde{t}_n]_{(2N+1)\times 1}$ which can be easily compared against $\bar{\boldsymbol{T}}$. Nevertheless, the $\tilde{\boldsymbol{\sigma}}_x$ and $\tilde{\boldsymbol{\sigma}}_y$ matrices cannot be extracted by comparing $\bar{\boldsymbol{T}}$, $\bar{\boldsymbol{R}}$ against $\boldsymbol{T}$, $\boldsymbol{R}$ since such an extraction would necessitate inversion of a non-invertible matrix. This issue can be resolved by comparing the continuity condition of the tangential electric fields in the original structure against that of the proposed model, i.e. (7) and (30), which necessitates the following relations between $\bar{\boldsymbol{T}}$, $\bar{\boldsymbol{B}}$ and $t_0$, $b_0$:

$$\bar{\boldsymbol{T}}=\mathbf{M}t_0,\quad \bar{\boldsymbol{B}}=\mathbf{M}b_0, \tag{45a}$$

where

$$[\mathbf{M}]_{(2N+1)\times 1}=\bar{\mathbf{Z}}^{(\mathrm{II})^{-1}}\boldsymbol{P}\eta_0\kappa \tag{45b}$$

Therefore, $\bar{\boldsymbol{T}}$ and $\bar{\boldsymbol{B}}$ in (31) can be written in terms of $t_0$ and $b_0$ to obtain the relation between $\bar{\boldsymbol{R}}$ and $t_0$. The obtained expression can then be easily compared against the similar expression between $\boldsymbol{R}$ and $t_0$ given in (12) and (13). In this fashion, the sought-after vector $\bar{\sigma}_x=(\sigma_x^{-N},\ldots,\sigma_x^{0},\ldots\sigma_x^{N})^t$, which contains the 2$N$+1 Fourier expansion coefficients of the surface conductivity can be written as follows:

$$\bar{\sigma}_x=\frac{\boldsymbol{\Gamma}^{-1}}{[\boldsymbol{S}_{21}\boldsymbol{A}+(\boldsymbol{S}_{22}-1)b_0]}\times[(\boldsymbol{I}+\boldsymbol{S}_{11}-\mathbf{M}\boldsymbol{S}_{21})\boldsymbol{A}+\ (-\mathbf{M}(\boldsymbol{S}_{22}+1)+\boldsymbol{S}_{12})b_0] \tag{46}$$

where, $\boldsymbol{\Gamma}=[\Gamma_{ij}]_{(2N+1)\times(2N+1)}$ is defined as

$$\Gamma_{ij}=\begin{cases}\eta_0\kappa p^{+}_{i+j-2(N+1)} & |j+j-2(N+1)|\le N\\ 0 & o.w.\end{cases} \tag{47}$$

Now, the fact that the amplitudes of the incident waves, i.e. $\boldsymbol{A}$ and $b_0$, appear in the mathematical expression of the surface conductivity seems surprising since the dependence of the surface conductivity on the incident wave amplitudes leads to the wrong belief that the proposed model is nonlinear. It should be, however, noted that the proposed model is not nonlinear and merely depends on the height of slits in the array. This point is clearly demonstrated below. In case the structure is semi-infinite, the backward waves $\bar{\boldsymbol{B}}$ and $b_0$ are zero and $\boldsymbol{A}=[a_n]_{(2N+1)\times 1}$, $a_n=|\alpha|\delta_{(N+1)0}$. Thus, (46) can be further simplified and written as

$$\bar{\sigma}_x = \frac{\mathbf{\Gamma}^{-1}}{\boldsymbol{S}_{21}\boldsymbol{U}} \times (\boldsymbol{I} + \boldsymbol{S}_{11} - \mathbf{M}\boldsymbol{S}_{21})\boldsymbol{U} \quad (48)$$

where $\boldsymbol{U} = [u_n]_{(2N+1)\times 1}, u_n = \delta_{(N+1)0}$. Evidently, the Fourier expansion coefficients of the surface conductivity do not depend on the amplitudes of the incident waves.

In case the height of slits in the array is finite and is equal to *h*, the vectors containing the 2*N*+1 Fourier expansion coefficients of the periodic surface conductivity created at the upper and lower interfaces of the structure, i.e. $\bar{\sigma}_x^u$ and $\bar{\sigma}_x^d$, can be easily obtained by simplifying (46):

$$\bar{\sigma}_x^u = \frac{\mathbf{\Gamma}^{-1}}{[\boldsymbol{S}_{21} + (\boldsymbol{S}_{22} - 1)\dfrac{\boldsymbol{S}_{22}\boldsymbol{S}_{21}}{e^{2j\beta_0 h} - \boldsymbol{S}_{22}^2}]\boldsymbol{U}} \times [(\boldsymbol{I} + \boldsymbol{S}_{11} - \mathbf{M}\boldsymbol{S}_{21}) + (-\mathbf{M}(\boldsymbol{S}_{22} + 1) + \boldsymbol{S}_{12})\frac{\boldsymbol{S}_{22}\boldsymbol{S}_{21}}{e^{2j\beta_0 h} - \boldsymbol{S}_{22}^2}]\boldsymbol{U} \quad (49a)$$

$$\bar{\sigma}_x^d = \frac{\mathbf{\Gamma}^{-1}}{\boldsymbol{S}_{22} - 1} \times (-\mathbf{M}(\boldsymbol{S}_{22} + 1) + \boldsymbol{S}_{12}) \quad (49b)$$

Once again, it can be easily seen that the Fourier expansion coefficients of the surface conductivity do not depend on the amplitude of the incident waves.

As for the bulk parameters of the effective model, the expressions for the permittivity and permeability components do not differ from what has been previously derived in [23]:

$$\varepsilon_x = \frac{1}{|p_0^+|^2}, \quad \mu_y = |p_0^+|^2, \quad \varepsilon_z = \infty \quad (50)$$

where $p_0^+$ is given in (11).

It is worth noting that the plasma-like behavior of 1D metallic grating is not on account of negative values for the effective permittivity rather it stems from the folding of the band structure caused by the periodicity of the surface conductivity.

*B. TE Polarization*

In a similar fashion, the continuity condition of the tangential electric fields in the original structure can be compared against that of the proposed model, i.e. (19) and (40), to obtain the following relations between $\bar{\boldsymbol{T}}$, $\bar{\boldsymbol{B}}$ and $t_0$, $b_0$:

$$\bar{\boldsymbol{T}} = \mathbf{N}t_0, \quad \bar{\boldsymbol{B}} = \mathbf{N}b_0, \quad (51a)$$

where

$$[\mathbf{N}]_{(2N+1)\times 1} = \boldsymbol{Q\kappa} \quad (51b)$$

By writing $\bar{\boldsymbol{T}}$ and $\bar{\boldsymbol{B}}$ in (41) in terms of $t_0$ and $b_0$ and then comparing the resultant equation against the similar equation between $\boldsymbol{R}$ and $t_0$ given in (12) and (24), the sought-after vector $\bar{\sigma}_y = (\sigma_y^{-N}, \ldots, \sigma_y^0, \ldots \sigma_y^N)^t$ containing 2*N*+1 Fourier expansion coefficients of the surface conductivity is obtained as follows:

$$\bar{\sigma}_y = \frac{\mathbf{\Omega}^{-1}}{[\boldsymbol{S}_{21}\boldsymbol{A} + (\boldsymbol{S}_{22} + 1)b_0]} \times [(\boldsymbol{Y}^{(\mathrm{I})}(\boldsymbol{I} - \boldsymbol{S}_{11}) - \bar{\boldsymbol{Y}}^{(\mathrm{II})}\mathbf{N}\boldsymbol{S}_{21})\boldsymbol{A} + (\bar{\boldsymbol{Y}}^{(\mathrm{II})}\mathbf{N}(1 - \boldsymbol{S}_{22}) - \boldsymbol{Y}^{(\mathrm{I})}\boldsymbol{S}_{12})b_0] \quad (52)$$

where $\mathbf{\Omega} = [\Omega_{ij}]_{(2N+1)\times(2N+1)}$ is defined as

$$\Omega_{ij} = \begin{cases} \kappa q^+_{i+j-2(N+1)} & |j + j - 2(N+1)| \le N \\ 0 & o.w. \end{cases} \quad (53)$$

For the semi-infinite structure where $\boldsymbol{A} = [a_n]_{(2N+1)\times 1}, a_n = |\alpha|\delta_{(N+1)0}$ and $b_0 = 0$, (52) can be written in a more simplified form

$$\bar{\sigma}_y = \frac{\mathbf{\Omega}^{-1}}{\boldsymbol{S}_{21}\boldsymbol{U}} \times [\boldsymbol{Y}^{(\mathrm{I})}(\boldsymbol{I}-\boldsymbol{S}_{11}) - \bar{\boldsymbol{Y}}^{(\mathrm{II})}\mathbf{N}\boldsymbol{S}_{21}]\boldsymbol{U} \tag{54}$$

For the structure with a finite height $h$, the vectors containing the $2N+1$ Fourier expansion coefficients of the periodic surface conductivity created at the upper and lower interfaces of the structure, i.e. $\bar{\sigma}_y^u$ and $\bar{\sigma}_y^d$, are easily obtained using (52):

$$\bar{\sigma}_y^u = \frac{\mathbf{\Omega}^{-1}}{[\boldsymbol{S}_{21} + (\boldsymbol{S}_{22}+1)\frac{\boldsymbol{S}_{22}\boldsymbol{S}_{21}}{e^{2j\beta_1 h} - \boldsymbol{S}_{22}^2}]\boldsymbol{U}} \times [(\boldsymbol{Y}^{(\mathrm{I})}(\boldsymbol{I}-\boldsymbol{S}_{11}) - \bar{\boldsymbol{Y}}^{(\mathrm{II})}\mathbf{N}\boldsymbol{S}_{21}) + (\bar{\boldsymbol{Y}}^{(\mathrm{II})}\mathbf{N}(1-\boldsymbol{S}_{22}) - \boldsymbol{Y}^{(\mathrm{I})}\boldsymbol{S}_{12})\frac{\boldsymbol{S}_{22}\boldsymbol{S}_{21}}{e^{2j\beta_1 h} - \boldsymbol{S}_{22}^2}]\boldsymbol{U} \tag{55a}$$

$$\bar{\sigma}_y^d = \frac{\mathbf{\Omega}^{-1}}{\boldsymbol{S}_{22}+1} \times [\bar{\boldsymbol{Y}}^{(\mathrm{II})}\mathbf{N}(1-\boldsymbol{S}_{22}) - \boldsymbol{Y}^{(\mathrm{I})}\boldsymbol{S}_{12}] \tag{55b}$$

As for the bulk parameters of the effective model, the expressions for the permittivity and permeability components do not differ from what has been previously derived in [2]:

$$\varepsilon_y = \frac{1}{2|q_0^+|^2}(1-\frac{\omega_p^2}{\omega^2}), \quad \mu_x = 2|q_0^+|^2, \quad \mu_z = \infty \tag{56}$$

where $q_0^+$ is given in (23) and $\omega_p = \pi c/a$. Interestingly, the plasma like behavior of the structure can be attributed to the negative values of $\varepsilon_y$ observed at frequencies lower than $\omega_p$.

## V. Numerical Results

In order to evaluate the accuracy of the proposed effective medium model, several numerical examples are examined. Full-wave simulations are performed with the commercial simulator COMSOL Multiphysics. In all simulations, metals are considered as perfect electric conductors. As the first numerical example, we suppose that the structure is illuminated by a TM polarized plane wave. The geometrical parameters of the structure are $d = 200\,\mu\mathrm{m}$, $h = 300\,\mu\mathrm{m}$ and $a = 50\,\mu\mathrm{m}$. The diffraction efficiencies [29] of the 0th and -1st transmitted order are plotted in Fig. 3 versus the normalized frequency $d/\lambda$ and the incident angle $\theta$. Also, diffraction efficiencies of the -2nd and +1st transmitted order versus the normalized frequency and the incident angle for the previous example are plotted in Fig. 4.

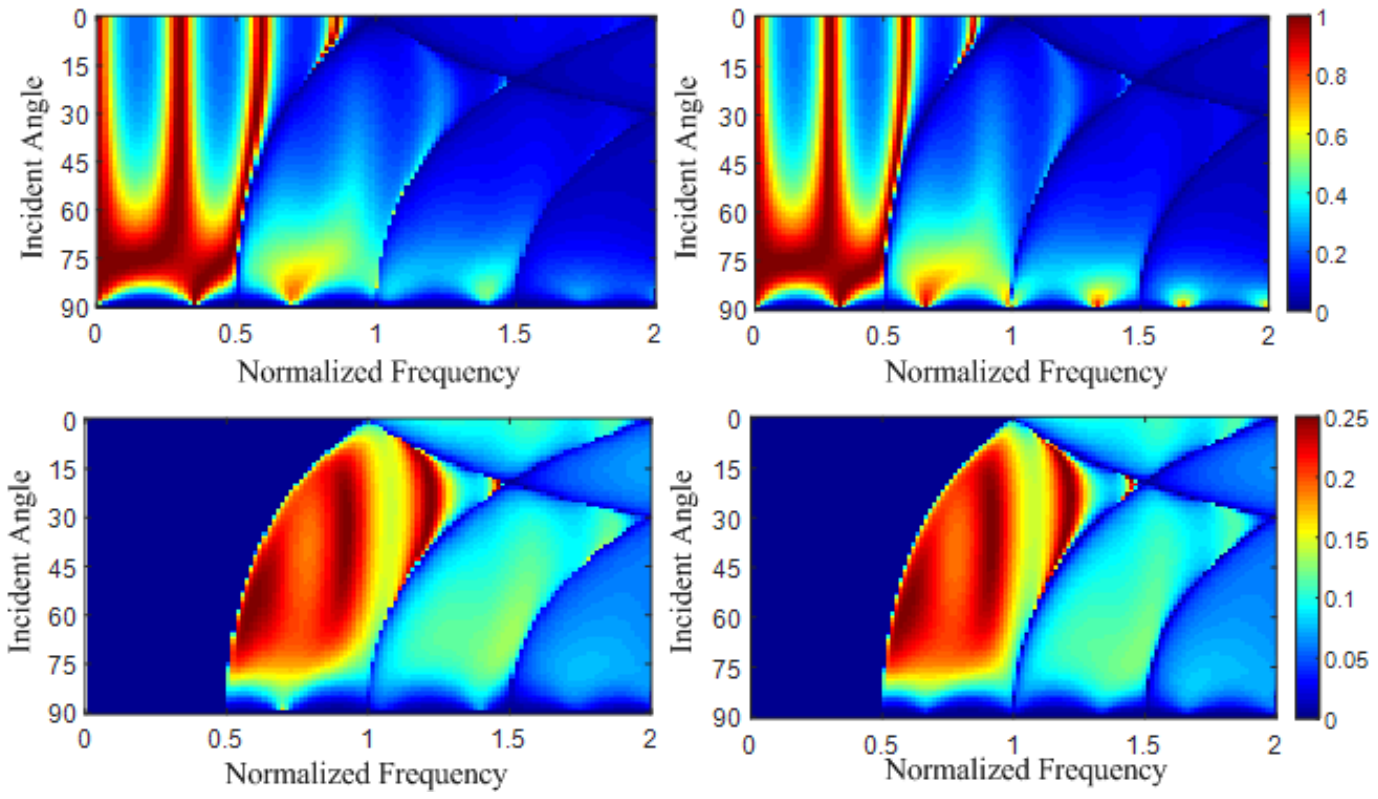

Fig. 3 The diffraction efficiencies of the 0th (top row) and -1st transmitted order (bottom row) for the TM incident wave obtained by full-wave simulations (left panel) as well as the effective medium model (right panel).

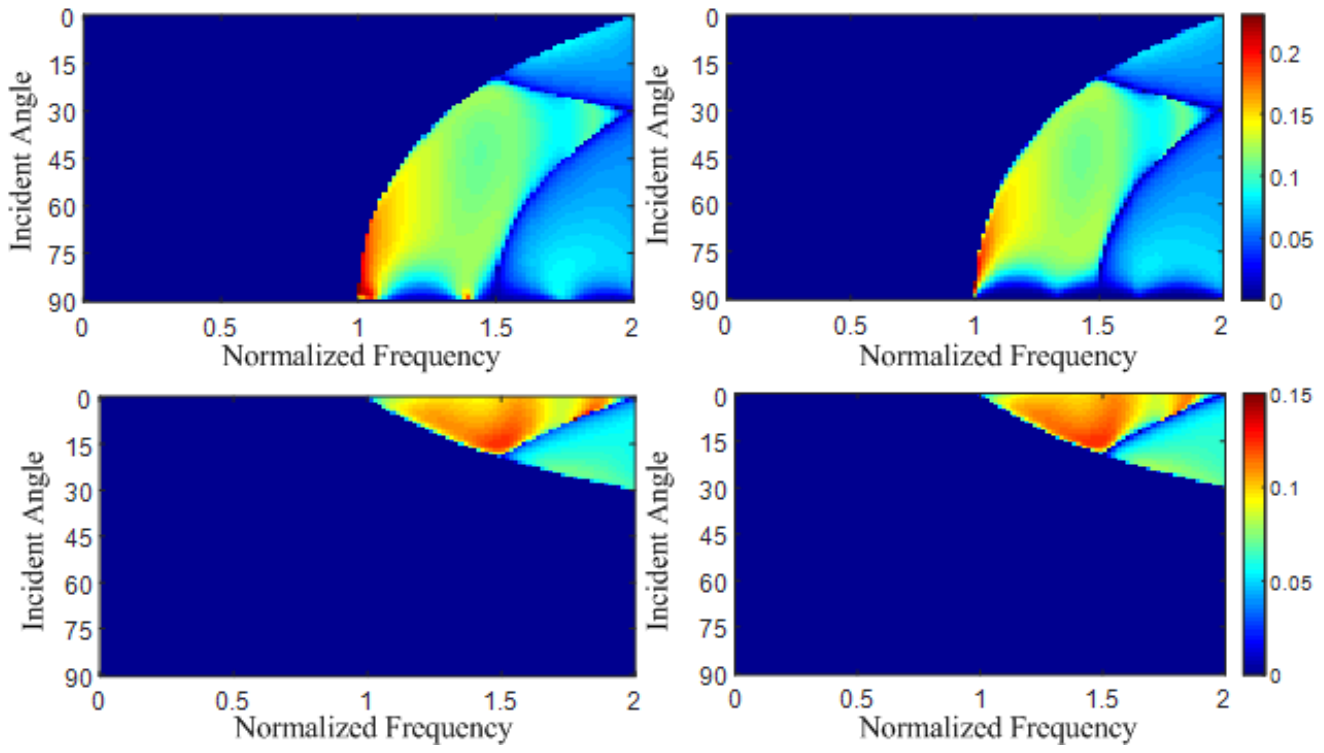

Fig. 4 The diffraction efficiencies of the -2nd (top row) and +1st (bottom row) transmitted order for the TM incident wave obtained by full-wave simulations (left panel) as well as the effective medium model (right panel).

Figs. 3 and 4 clearly demonstrate an excellent agreement between the results obtained by the full-wave simulations (left panels) and the effective medium model (right panels). It should be noticed that we have kept five diffraction orders (*N*=2) in the derivation of the periodic surface conductivity expression in (49) and thus only the 0th-, $\pm 1$ st- and $\pm 2$ nd-order diffracted waves are meticulously modeled by the proposed approach.

Next, we demonstrate the accuracy of the proposed method in modeling the sharp Fano-type EOT resonance of the structure. The 0th-order transmission coefficient versus the normalized frequency for the structure of Fig. 3 with parameters $d = 200\mu\text{m}$, $h = 300\mu\text{m}$ and $a = 50\mu\text{m}$ when illuminated by a TM polarized plane wave is plotted in Fig. 5. The results are obtained by using full-wave simulations (solid line) and the effective medium model (dashed line). Evidently, there is an excellent agreement between the two methods.

Fano resonances have been extensively studied in periodic structures [28]. The asymmetric line shape of Fano resonance is attributed to the coupling between the propagating continuum states and the evanescent bound states. As our model is capable to accurately predict all diffraction orders in the structure, it can be used to study the Fano-type EOT resonance in terms of coupling between near cut-off Floquet orders.

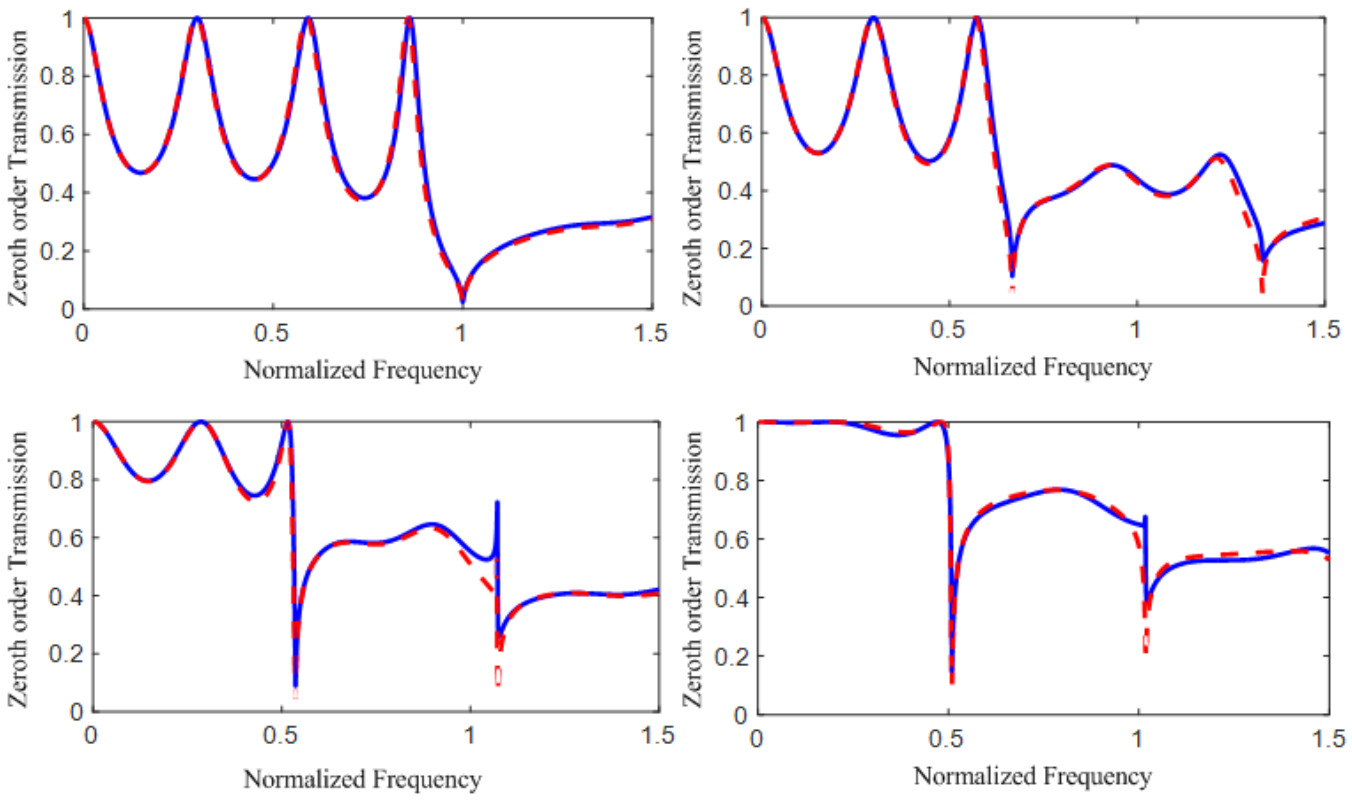

Fig. 5 The magnitude of zeroth-order transmission coefficient obtained by full-wave simulations (solid line) and effective medium model (dashed line) for incident angles $0^\circ$ (top left), (b) $30^\circ$ (top right), (c) $60^\circ$ (bottom left) and (d) $75^\circ$ (bottom right).

Next, we plot the dispersion diagram of the first three even waveguide modes of the metallic grating (solid line) and its corresponding effective isotropic (dots) and anisotropic (dashed line) models proposed in [2] for the TM polarization. The geometrical parameters of the original structure are $d = 200\mu\text{m}$, $h = 1250\mu\text{m}$ and $a = 50\mu\text{m}$. The results are shown in Fig. 6. Apparently, the plasma-like behavior of the 1D metallic grating which translates into the existence of bound surface waves with zero group velocity cannot be understood by resorting to the positive values of the effective permittivity and permeability of the structure. As a point of fact, the dispersion diagram obtained by the homogenized isotropic or anisotropic models does not significantly differ from the dispersion diagram of the original structure with deep-subwavelength slits. Still, there is a conspicuous discrepancy between the dispersion diagram of the isotropic and anisotropic waveguide models and that of the 1D metallic slit arrays when the factor $a/d$ is large [2]. In such cases, it is necessary to resort to periodic models that can mimic the behavior of the original structure. The proposed model which has a periodic surface conductivity profile is an example of such models, which can attribute the slow light behavior of the structure to the folding of the band diagram.

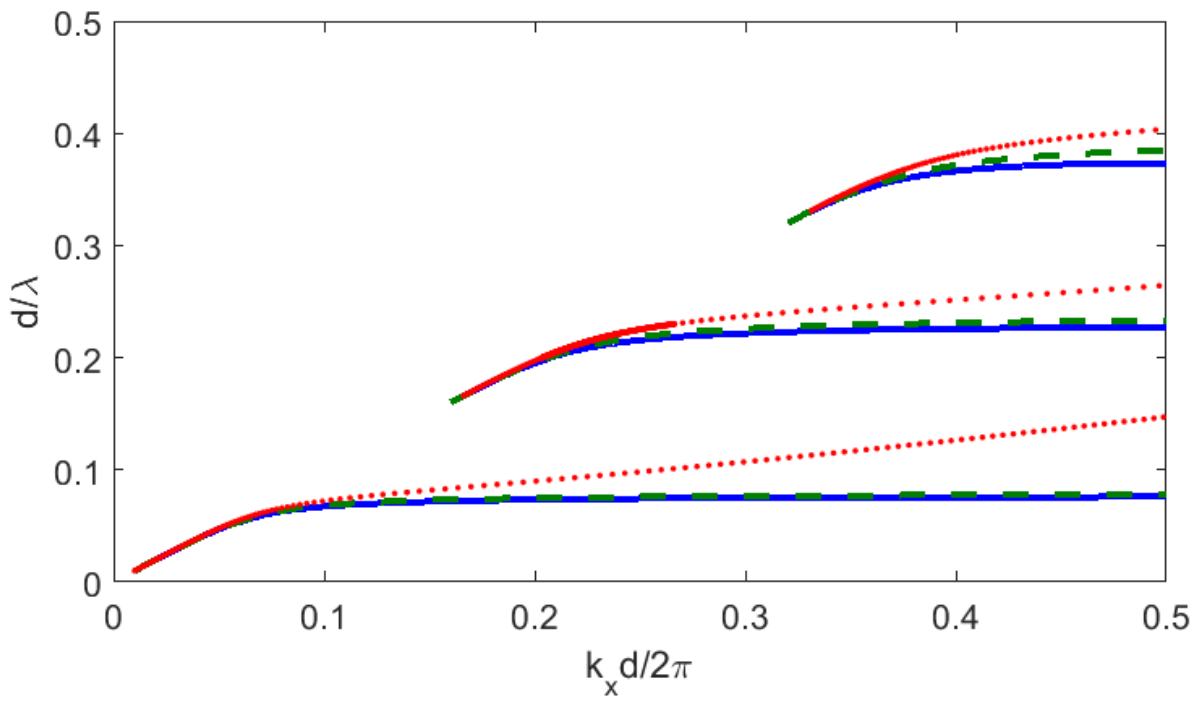

Fig. 6 The dispersion diagram of the first three even modes of the metallic grating (solid line) and its corresponding effective isotropic (dots) and anisotropic (dashed line) models proposed in [2].

Finally, the accuracy of the proposed model for the TE polarized plane waves impinging on the structure is assessed. Fig. 7 shows the diffraction efficiencies of the 0th and -1st transmitted order versus the normalized frequency and the incident angle in the structure with parameters $d = 200\mu\text{m}$, $h = 600\mu\text{m}$ and $a = 165\mu\text{m}$. Once again, an excellent agreement is observed between our results and numerical ones.

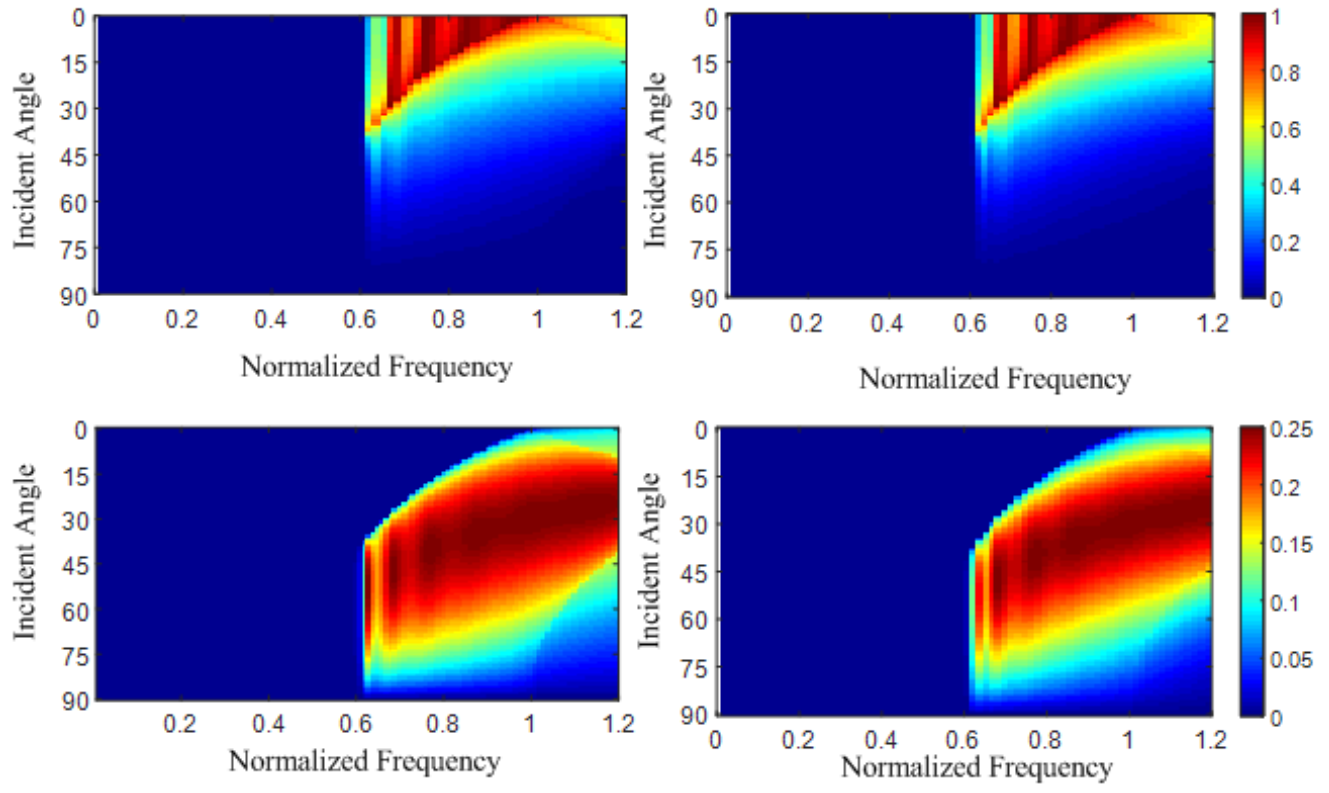

Fig. 7 The diffraction efficiencies of the 0th (top row) and -1st (bottom row) transmitted order for the TE incident wave obtained by full-wave simulations (left panel) as well as the effective medium model (right panel)

## VI. Conclusion

In this work, we proposed an effective medium model for the one-dimensional periodic arrangement of cut-through slits in the microwave and terahertz regions where metals can be approximated by perfect electric conductors. The proposed analytical model consists of an anisotropic bulk medium and a periodic surface conductivity at the interface of the metallic grating and the surrounding media. The parameters of the effective medium model are given by explicit analytical expressions for both major polarizations and for all incident angles. This model is derived assuming only the principal mode is supported by the slits. The previously-proposed models are only capable of mimicking the zeroth-order scattered waves; whereas the here-proposed approach can meticulously emulate all diffraction orders by introducing the periodic surface conductivity in the effective medium model.

This theory opens up a new methodology for the analysis of more complex structures such as compound gratings and metamaterials designed to tailor the characteristics of higher Floquet orders. It can be also used to show that the plasma-like behavior of the structure in the TM polarization is due to the band folding of the structure. Additionally, the here-proposed fully analytical effective medium model may be utilized to further investigate the Fano-type EOT resonances and their application as Fano filters.